\documentclass{PoS}
\usepackage{breakurl}

\def \degree {$^{\circ}$}

\def\arxiv#1{\href{http://arxiv.org/abs/#1}{{\tt arXiv:#1}}}
\def\astroph#1{\href{http://arxiv.org/abs/astro-ph/#1}{{\tt astro-ph:#1}}}

\let\OLDthebibliography\thebibliography
\renewcommand\thebibliography[1]{
  \OLDthebibliography{#1}
  \setlength{\parskip}{0pt}
  \setlength{\itemsep}{0pt plus 0.3ex}
}

\title{ Searching for TeV gamma-ray emission associated with IceCube high-energy neutrinos using VERITAS}

\ShortTitle{VERITAS observations of IceCube HE neutrino positions}

\author{\speaker{M.~Santander},$^a$ for the VERITAS$^{\dagger}$ and IceCube$^{\star}$ Collaborations \\
\llap{$^a$} Department of Physics and Astronomy \\
Barnard College, Columbia University, New York, NY, USA \\
E-mail: \email{santander@nevis.columbia.edu} \\
\llap{$^{\dagger}$} http://veritas.sao.arizona.edu \\
\llap{$^{\star}$} http://icecube.wisc.edu/collaboration/authors/icrc15\_icecube
}

\abstract{A clue to finding the long-sought sources of cosmic rays is the recent observation of an 
astrophysical flux of high-energy neutrinos by the IceCube detector, since these possibly originate in hadronic 
interactions at cosmic-ray accelerators. While the neutrino sky map shows no significant indication of point sources so 
far, it is possible to utilize the sensitivity of Imaging Air Cherenkov Telescope (IACT) arrays, such as VERITAS, to search for hadronic 
gamma-ray emission from the neutrino directions.

Over the last 2 years, the positions of neutrino events detected by IceCube have been observed using the
 VERITAS array. Observations have been limited to muon neutrino events, since their typical angular reconstruction 
 uncertainty is below 1$^{\circ}$, smaller than the 3.5$^{\circ}$ diameter of the VERITAS field of view. The 
 location of VERITAS further constrains the neutrino event positions that can be observed to those located in 
 the northern sky, or at moderate southern declinations. The list of observed positions was selected from published 
 results and a set of high-energy muon tracks provided by IceCube. We present the current status and some 
 preliminary results from this program.}

\FullConference{The 34th International Cosmic Ray Conference,\\
		30 July- 6 August, 2015\\
		The Hague, The Netherlands}

\begin{document}

\section{Introduction}

The discovery of an astrophysical flux of high-energy neutrinos by IceCube \cite{HESE1, HESE2} is a major step 
forward in the ongoing search for the origin of cosmic rays, since the neutrino emission may be produced
by hadronic interactions in astrophysical accelerators. 

IceCube's detection was based on an analysis that searched for energetic events that have their interaction vertex 
contained inside the detector volume. The latest update of this analysis~\cite{HESE2} has revealed 37 \textit{contained} 
events in the TeV--PeV energy range observed over the course of three years, out of which 28 are particle cascades 
(produced by charged interactions of $\nu_{e, \tau}$ or neutral interactions of any neutrino flavor) with angular resolutions 
of about 15\degree~and one was determined to be a background event. The event topology of the remaining 8 is compatible with $\nu_{\mu}$-induced muon 
tracks with an angular resolution of about 1\degree. 

The astrophysical spectrum derived from this analysis is consistent with a $E_{\nu}^{-\Gamma}$ power-law with 
$\Gamma = 2$ and a normalization of $E^2_{\nu} \Phi_{\nu}(E_{\nu}) = 1.2 \times 10^{-8} $ GeV / cm$^{2}$ s sr per 
neutrino flavor. Additional IceCube analyses lowering the energy threshold seem to favor softer spectra, with 
$\Gamma$ in the 2.3 -- 2.5 range~\cite{Jakob}. A separate analysis looking only at up-going \textit{uncontained} 
muon tracks (i.e. their vertex is allowed to be located outside the detector volume) also found evidence at the 
$4\sigma$ level for a diffuse astrophysical flux component with similar normalization and spectrum. This is an 
important analysis for gamma-ray follow-ups since the removal of the containment constraint for up-going muons 
provides a large sample of events ($\sim$ 10 per year) with good angular resolution ($<$ 1\degree) compared to 
the $\sim 3$ muon events per year detected using the contained analysis. In this work we make use of these muon events to search for gamma-ray counterparts.

The sky distribution of the 37 contained events 
does not show a significant deviation from isotropy. The most significant excess is located near the Galactic Center  (p-value: 8\%.) The search for neutrino point sources using uncontained muons does not reveal any significant hotspots and 
has been used to set upper limits on the neutrino flux from point sources that are at the $10^{-2}$ level of the all-sky 
astrophysical diffuse flux~\cite{HESE2}. For steady sources, this implies a large number of them ($N_s > 10^2$) distributed across 
the sky since otherwise the point-source bound would be exceeded. See Ref.~\cite{ICnuReview} for a review of the 
astrophysical implications of the IceCube result.

The lack of correlation with the galactic plane and the apparent isotropy of the neutrino data set has favored a 
number of extragalactic interpretations. This represents a challenge for the detection of pionic gamma rays associated with the neutrino events since they are quickly attenuated by the Extragalactic Background Light (EBL).  For instance, the attenuation length of a 1 PeV gamma ray is similar to the distance to the Galactic Center. 
However, if the hadronic gamma-ray emission spans several orders of magnitude in energy, as is expected in a Fermi acceleration process, it can be extrapolated down to lower energies (GeV to TeV) where the EBL attenuation is not as severe. So far, the most distant gamma-ray source detected in the very-high-energy range (VHE, E > 100 GeV) is the gravitationally-lensed blazar  S3 0218+37 at z = 0.944 observed during a flare with the MAGIC telescopes~\footnote{\texttt{http://www.astronomerstelegram.org/?read=6349}}.

\begin{figure}[h]
\centering
\includegraphics[width=0.6\textwidth]{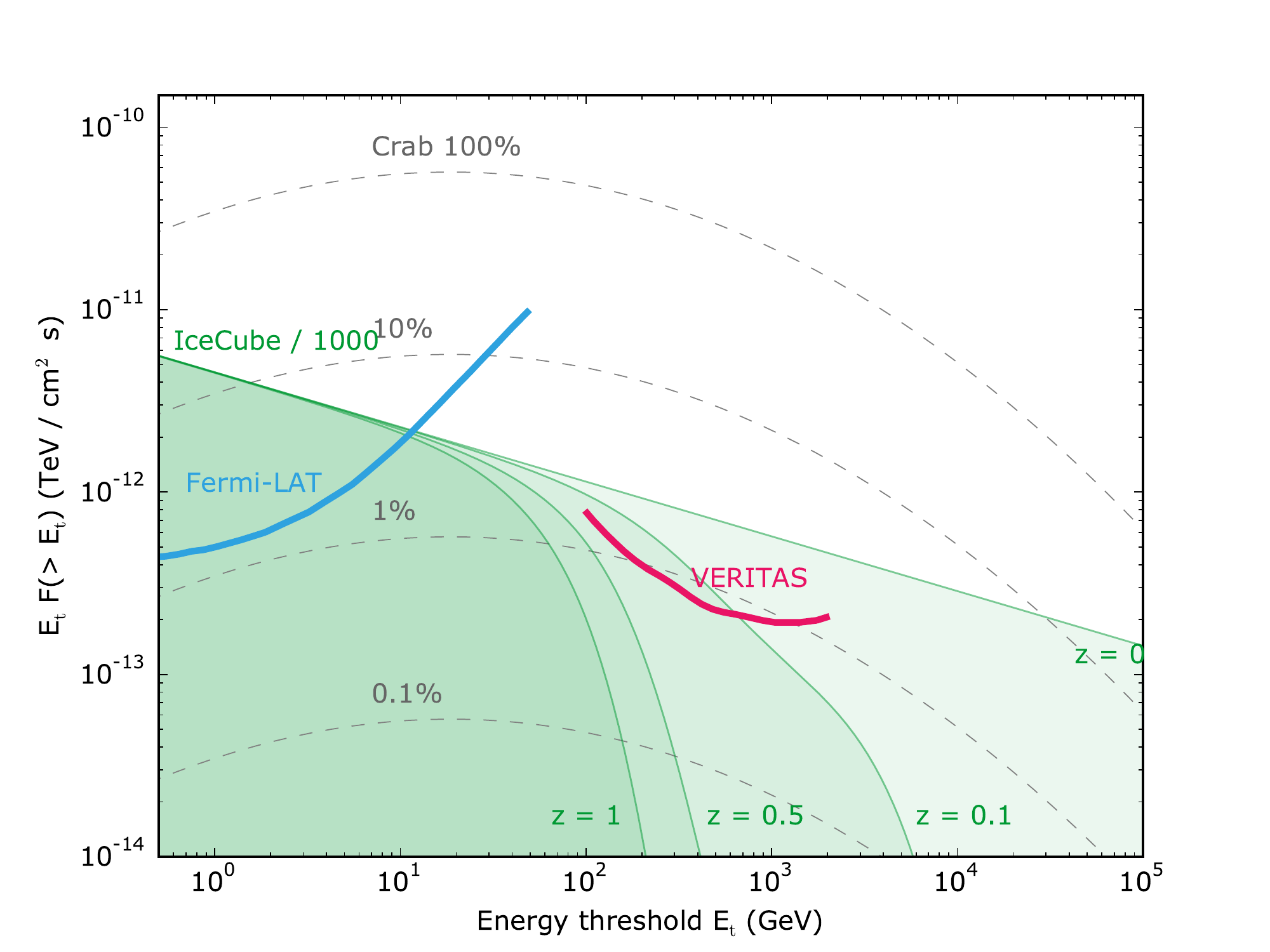}
\caption{Extrapolation to lower energies of the IceCube neutrino flux converted to gamma rays (1 GeV to 100 TeV) compared with the integral sensitivity $E_t \; F(E > E_t)$ of VERITAS and \emph{Fermi}-LAT above the energy threshold $E_t$. For VERITAS~\cite{Hillas} the $5\sigma$ sensitivity is given for a 50-hour exposure, while for the \emph{Fermi}-LAT (extragalactic)~\cite{Fermi} the sensitivity is for a 1 year exposure. The IceCube flux has been scaled by $10^{-3}$ and accounts for EBL absorption using the model by Franceschini et al.~\cite{EBL} for sources at different redshifts.}

\label{ic_spectrum}
\end{figure}

In Fig.~\ref{ic_spectrum} we show an extrapolation of the all-sky IceCube neutrino flux (obtained from \cite{HESE2} 
with $\Gamma=2.3$) which has been converted at a 1:1 ratio to gamma rays down to GeV energies. Detailed conversion procedures assuming p--p interactions can be found in the literature~\cite{Kappes}. The spectra have been scaled by a factor of $10^{-3}$ and account for EBL absorption using the model of Franceschini et al.~\cite{EBL} for source distances of z = 0, 0.1, 0.5, 1. Reprocessed gamma-rays from the pair-production and inverse Compton scattering of CMB photons are not considered in this illustration. 

As a comparison, the integral sensitivity of Fermi-LAT and VERITAS to steady point sources 
is shown. For VERITAS, this sensitivity represents the flux necessary above a given energy threshold to achieve 
a $5\sigma$ detection in 50 hours of observation. It can be inferred from the plot that if there are $N_s < 10^3$ steady sources located at $z < 0.1$ that contribute equally to the all-sky neutrino flux, VERITAS can detect them in 50 hours gamma emission should be detectable at the $5\sigma$ level with 50 hours of VERITAS observations. 

A null detection by VERITAS using about 2 hours of observation per target can set flux limits at the 1 -- 2 \% level of 
the Crab nebula that would indicate more distant, or more numerous and faint, sources. An analysis of Fermi-LAT data that relies on the extrapolation of the neutrino flux to GeV energies has shown no significant association between GeV gamma-ray sources and the IceCube events~\cite{FermiEvents}.


In this work we present the current status and preliminary results from an observational program carried out with VERITAS 
to search for VHE emission from the location of high-energy muon neutrinos detected by IceCube. 
This search is mostly sensitive to steady sources, or variable sources observed during a high state, since in many cases the VERITAS observations are performed many months after the detection of the neutrino events by IceCube. Transient sources that flare only once, such as gamma-ray bursts, can not be detected in this search. 
A realtime alert system is currently in development to improve the sensitivity to transient 
sources.

\section{Detectors and data sets}

VERITAS (the Very Energetic Radiation Imaging Telescope Array System) is a ground-based instrument for VHE gamma-ray astronomy with maximum sensitivity in the 80 GeV to 30 TeV range. It is located at the Fred Lawrence Whipple Observatory (FLWO) in southern Arizona (31 40N, 110 57W, 1.3km a.s.l.). The array consists of four 12-m optical telescopes each equipped with a camera containing 499 photomultiplier tubes (PMTs) covering a field of view of $3.5^{\circ}$. The cameras of the four VERITAS telescopes provide a stereoscopic view of gamma-ray showers in the atmosphere through the detection of the Cherenkov light they emit. The geometry and brightness of the shower images is used to reconstruct the energy and incoming direction of the primary gamma-ray photon.

\begin{figure}[h]
\centering
 \raisebox{-0.5\height}{\includegraphics[width=0.5\textwidth]{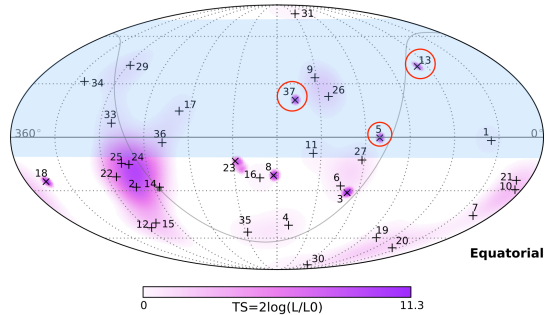}}
\caption{Sky map of contained high-energy neutrino events from~\cite{HESE2}. The fraction of the sky observable from the VERITAS site is shown in light blue. The positions of the IceCube contained muon tracks (C5, C13, and C37) has been highlighted with red circles. }
\label{integral}
\end{figure}

Over the last two years, VERITAS has been used to search for gamma-ray emission associated with high-energy neutrino events detected by IceCube. IceCube is a cubic-kilometer neutrino detector installed in the ice at the geographic South Pole~\cite{IC3} between depths of 1450\,m and 2450\,m. Detector construction started in 2005 and finished in 2010. Neutrino reconstruction relies on the optical detection of Cherenkov radiation emitted by secondary particles produced in neutrino interactions in the surrounding ice or the nearby bedrock. 

Only muon track positions have been observed with VERITAS since their angular uncertainty ($< 1^{\circ}$) is significantly smaller than the field of view of the VERITAS telescopes. The location of VERITAS in southern Arizona restricts the observable neutrino positions to those in the northern sky, or at low declinations in the southern sky.
The muon tracks observed under this program have been selected from two data sets: the list of contained events published by IceCube \cite{HESE2} that includes 3 muons in the northern sky (see Table~\ref{tab:muevents} for positions) and an unpublished list of 20 potentially astrophysical uncontained muon events that has been shared with VERITAS through a cooperation agreement. These uncontained events form the high-energy end of a sample of well-reconstructed muon tracks selected by IceCube to measure the astrophysical muon neutrino flux using exclusively the track channel. Details of this analysis will be available in a forthcoming publication~\cite{Chris}.

In this work, the three contained muons are identified with the letter `C'  followed by the event ID used in Ref.~\cite{HESE2} and the 20 uncontained muons have been labeled as `UC' followed by a correlative number. For each uncontained event, IceCube has provided:

\begin{itemize}
\setlength{\itemsep}{0pt}
\item \small{\textit{Celestial coordinates}: right ascension and declination.}
\item \textit{Muon energy}: to priorize higher energy events which have a higher chance of being astrophysical.
\item \textit{Angular uncertainty}: to discard events with poor angular resolution.
\end{itemize}

Further information about these 20 events is kept blind for the moment to prevent biasing other IceCube analyses.
The VERITAS observations were performed using the standard \emph{wobble} observation strategy where the telescopes are offset from the position of the potential source to allow for a simultaneous determination of the background. Offsets of $0.5^{\circ}$ and $0.7^{\circ}$ with respect to the best fit neutrino location were used to provide better coverage of the neutrino error circle. 

\begin{table*}[h]
\centering
\begin{tabular}{c | c | c | c
}
ID  & R.A. (deg.) & Decl. (deg.) & Energy (TeV) \\
\hline
C5  & 110.6$^{\circ}$ &  -0.4$^{\circ}$ & $71.4^{+9}_{-9}$ \\
C13  & 67.9$^{\circ}$ & 40.3$^{\circ}$ & $253^{+26}_{-22}$ \\
C37  & 167.3$^{\circ}$ & 20.7$^{\circ}$ & $30.8^{+3.3}_{-3.5}$ \\
\end{tabular}
\caption{High-energy muon event positions and energies (from \cite{HESE2}) observed with VERITAS.}
\label{tab:muevents}
\end{table*}

About 50\% of the observations for this program were conducted during moonless nights with good weather. A fraction of the data was taken during bright moonlight conditions with the cameras operating with reduced high voltage or when the weather was suboptimal. In those cases, the appropriate selection cuts have been applied to the data to remove periods with unstable trigger rates or high PMT currents in order to analyze only good quality data. Table~\ref{tab:uls} summarizes the observations of the IceCube targets.

\section{Analysis and preliminary results}

The analysis of VERITAS data involves introducing cuts to separate gamma-ray shower candidate events from a dominant background
of hadronic cosmic-ray showers. These cuts are applied on the parameters that characterize the geometry of the shower images. In this work, we have used \emph{soft} cuts optimized for sources with spectral indices of $\sim -4$. This choice is based on the assumption that neutrino sources are expected to be located at high-redshift and therefore will exhibit soft VHE spectra due to EBL absorption. 

\begin{figure}[h]
\centering
\raisebox{-0.5\height}{\includegraphics[width=0.55\textwidth]{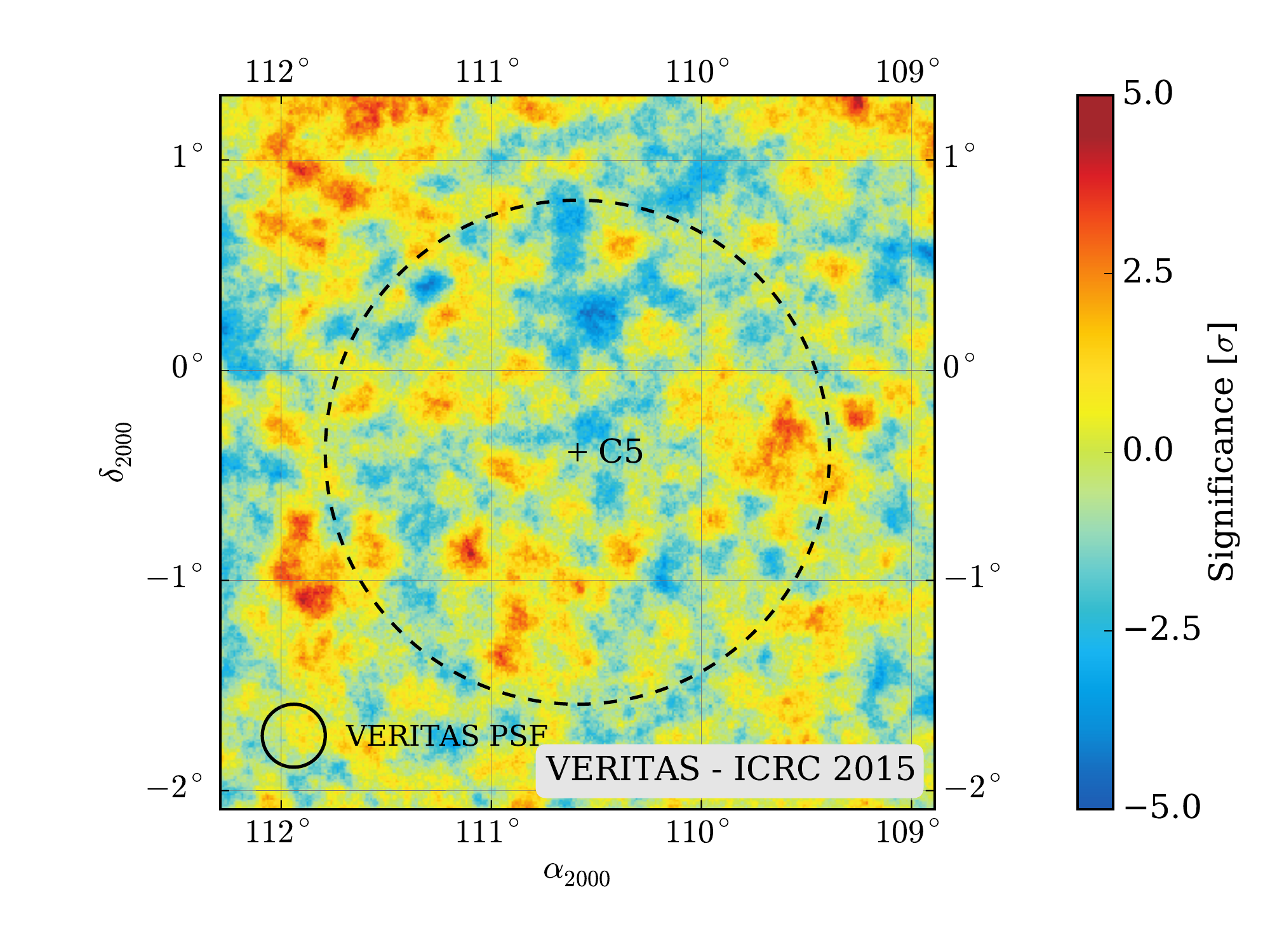}}
\raisebox{-0.5\height}{\includegraphics[width=0.44\textwidth]{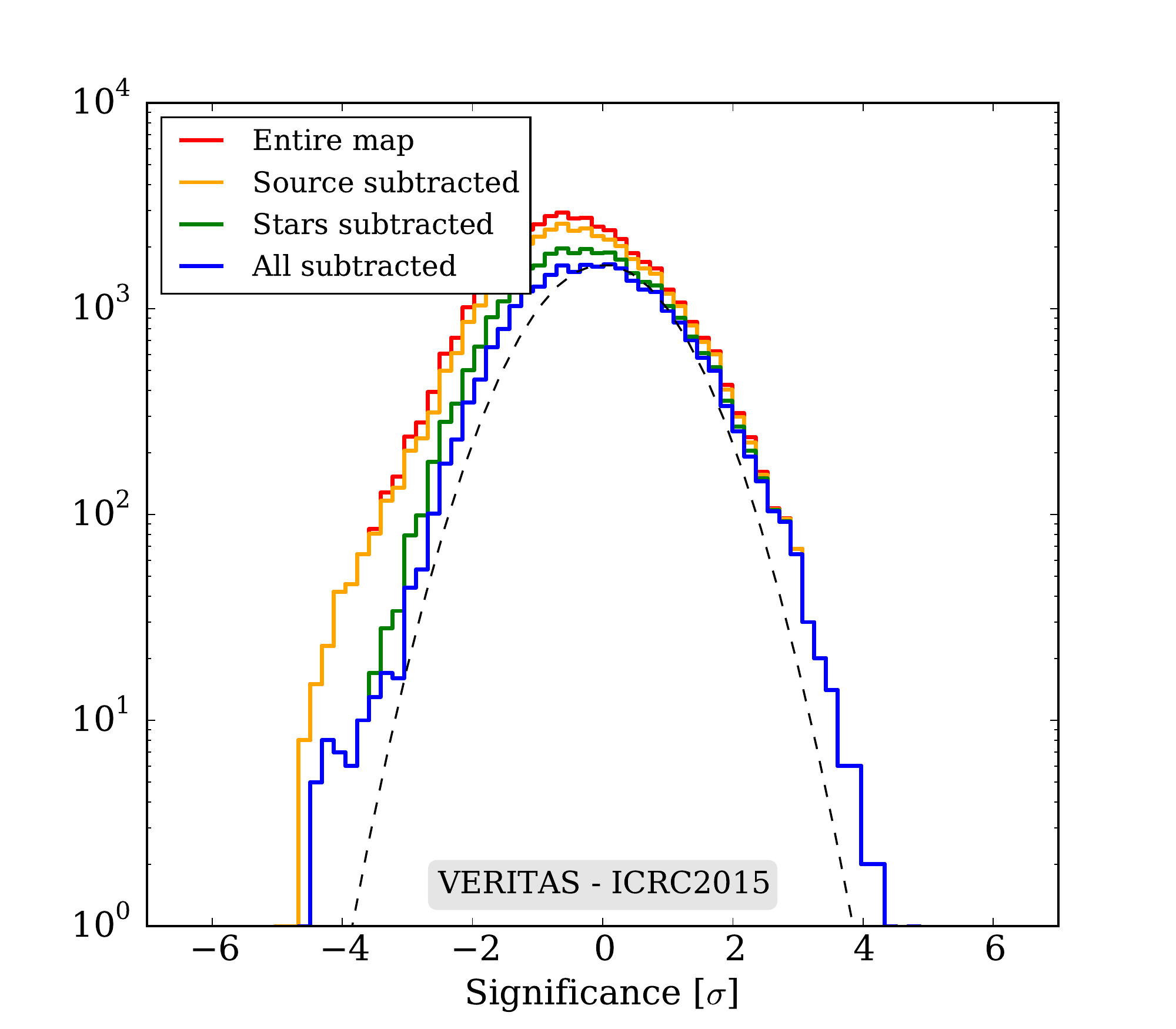}}
\raisebox{-0.5\height}{\includegraphics[width=0.55\textwidth]{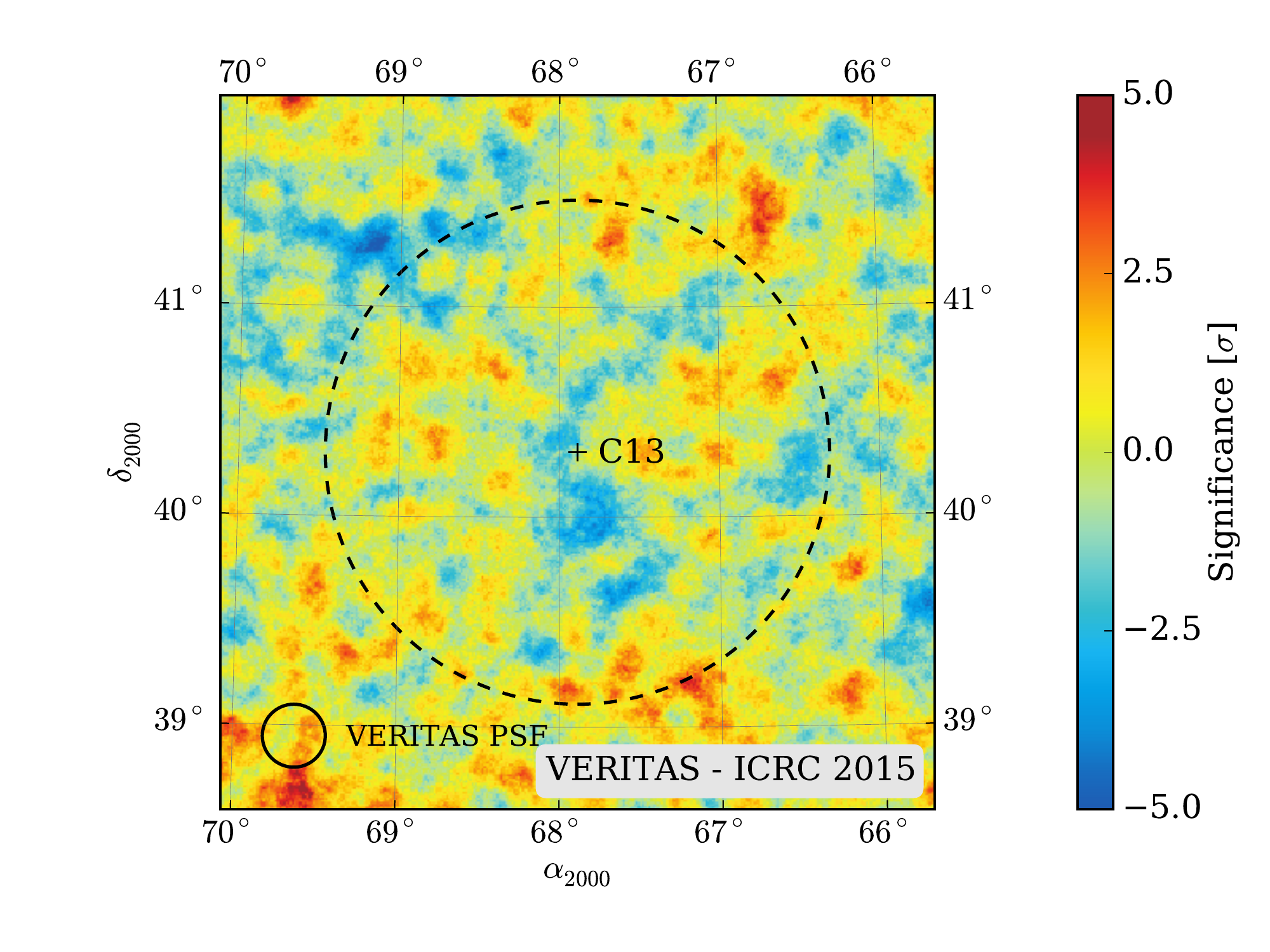}}
\raisebox{-0.5\height}{\includegraphics[width=0.44\textwidth]{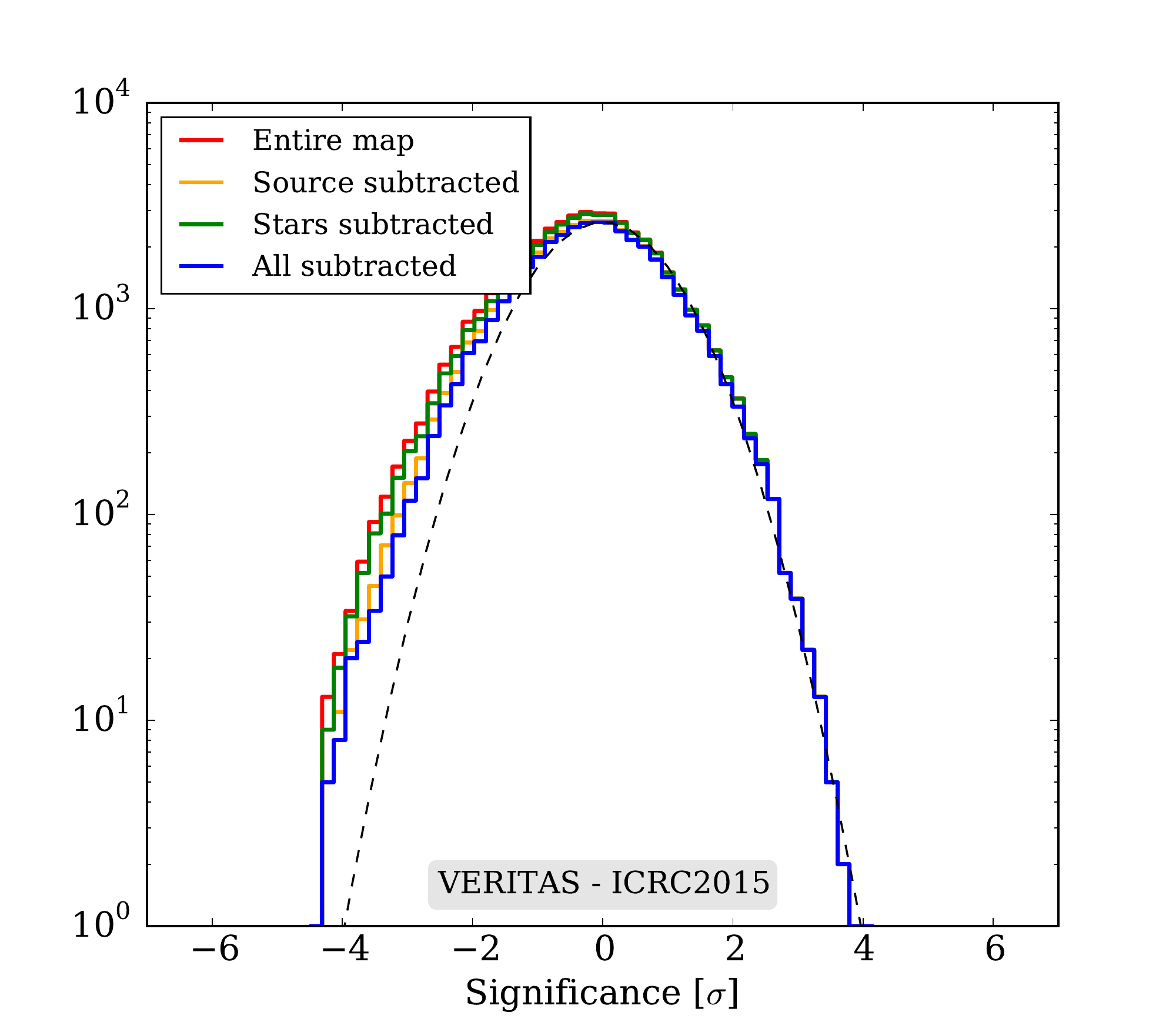}}
\raisebox{-0.5\height}{\includegraphics[width=0.55\textwidth]{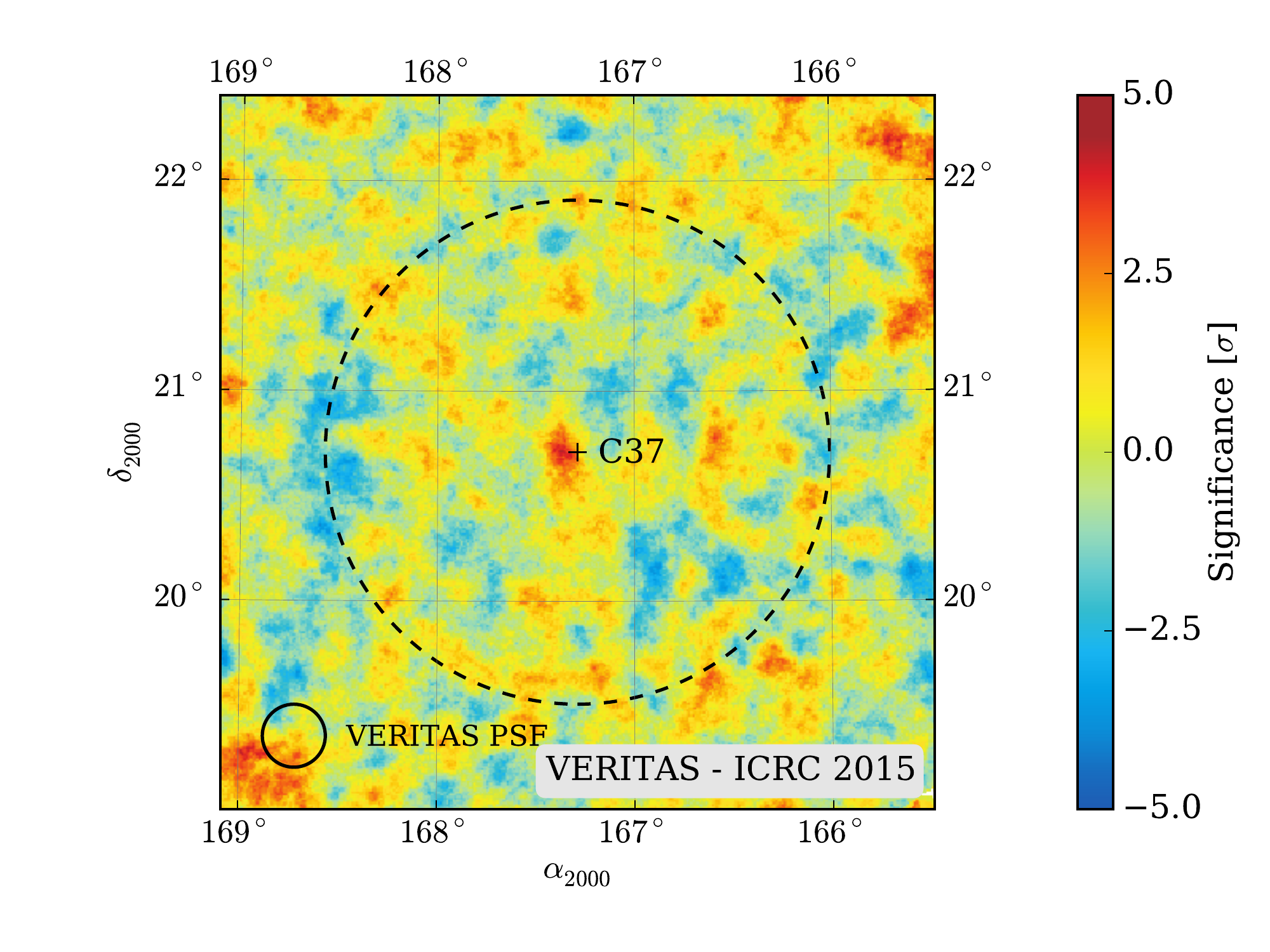}}
\raisebox{-0.5\height}{\includegraphics[width=0.44\textwidth]{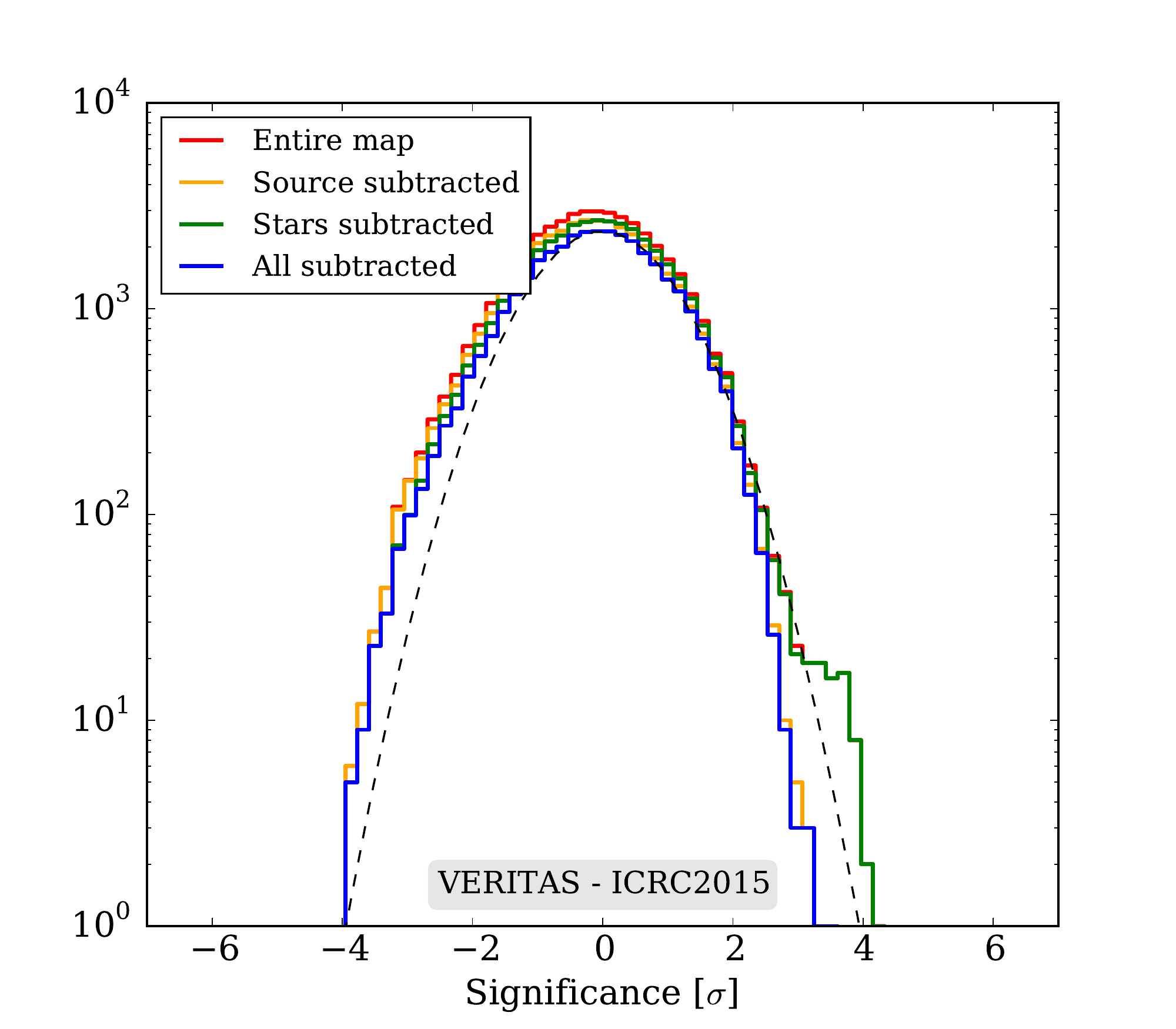}}
\caption{\emph{Left:} Preliminary significance maps for the three contained neutrino positions observed with VERITAS. From the top down: C5, C13, and C37. The dashed circle represents the angular uncertainty in the neutrino position as estimated by IceCube. The most significance excess in these maps is located near the best-fit position of the C37 track. We estimate that, after accounting for statistical trials, the significance of this excess is $< 2\sigma$. \emph{Right:} Significance histograms for the significance maps. No strong indication of a point source is evident given the lack of a positive excess in the Gaussian distributions.  }
\label{fig:signif}
\end{figure}

Preliminary results from the analysis of the gamma-ray data show no statistically significant excess in the regions near the neutrino positions. Sample significance maps are shown in Fig.~\ref{fig:signif} for the three contained muon tracks: C5, C13, and C37. The most significant excess in these maps is near the best-fit location of the track C37, with a pre-trials significance of $4.3\sigma$. We estimate that searching for an excess in each of these maps involves about 2000 trials, which corresponds to the ratio between the areas covered by the IceCube and VERITAS point-spread-functions. Taking trials into account, the significance of the excess is reduced to $\sim 2.0 \sigma$. Given the lack of a source detection, upper limits have been calculated on the gamma-ray flux at the best-fit neutrino locations. A list of track events and 99\% level pre-trials upper limits (ULs) are shown in Table~\ref{tab:uls} for gamma-ray energies above 100 GeV. 

Most ULs are at the level of a few percent of the gamma-ray flux of the Crab nebula, the traditional standard candle in VHE astrophysics. The astrophysical interpretation of these limits depends on a number of assumptions about the nature of the sources producing the neutrino flux. We assume that the astrophysical neutrinos are emitted by a large number of steady sources ($N_s$) that contribute equally to the total diffuse flux. If each of the neutrino events observed by VERITAS is astrophysical in origin(i.e. not a background muon), then the upper limits can be understood as a constraint on the fraction of the flux that each source contributes to the total all-sky flux. 

Taking Fig.~\ref{ic_spectrum} as a reference point, an upper limit at the level of a few percent of the Crab is compatible with each source contributing only one part-per-mille of the total flux or, put otherwise, that $N_s > 1000$ if all sources have the same flux with an spectral index of -2.3 and a 1:1 gamma-neutrino ratio. This is only valid if the sources are at low redshifts. For higher redshifts (z > 0.2) a smaller number of steady sources can't be ruled out since their gamma-ray flux is attenuated by the EBL. However, existing point-source upper limits from IceCube potentially indicate that $N_s > 100$ as already mentioned~\cite{HESE2}.
Future work will explore the implications of the upper limits on different types of classes with known redshift evolutions and the potential contribution of sources in our galaxy. This may help further constrain the density of astrophysical neutrino sources. 
 
\begin{table*}[h]
\centering
\begin{tabular}{c | c | c | c}
ID  & Observation time [min] & UL (99\%) [cm$^{-2}$ s$^{-1}$] & UL (99\%) [C.U.]\\
\hline
C5 & 180 & $8.33 \times 10^{-12}$ & 2.3\%\\
C13 & 574 & $4.01 \times 10^{-12}$ & 1.1\% \\
C37 & 275 & $7.30 \times 10^{-12}$ & 2.0\% \\
\hline
\hline
UC2 & 25 & $2.12 \times 10^{-11}$ & 5.8\% \\
UC3 & 180 & $6.31 \times 10^{-12}$ & 1.7\% \\
UC4 & 122 & $9.89 \times 10^{-12}$ & 2.7\% \\
UC5 & 90 &  $6.66 \times 10^{-12}$ & 1.8\% \\
UC6 & 25 & $9.53 \times 10^{-12}$ & 2.6\% \\
UC7 & 15 & $3.96 \times 10^{-11}$ & 10.9\% \\
UC8 & 60 & $9.31 \times 10^{-12}$ & 2.6\%\\
UC9 & 40 & $1.52 \times 10^{-11}$ & 4.2\%\\
UC10 & 90 & $9.40 \times 10^{-12}$ & 2.6\% \\
UC11 & 209 & $4.4 \times 10^{-12}$ & 1.2\% \\
UC12 & 25 & $9.53 \times 10^{-12}$ & 2.6\% \\
UC15 & 90 & $7.40 \times 10^{-12}$ & 2.0\% \\
UC16 & 40 & $8.57 \times 10^{-12}$ & 2.4\% \\
UC17 & 150 & $4.41 \times 10^{-12}$ & 1.2\%\\
UC19 & 210 & $3.92 \times 10^{-12}$ & 1.1\% \\
\end{tabular}
\caption{Summary of the observations of high-energy muons performed so far. For each contained (`C') and uncontained (`UC') IceCube event, the amount of VERITAS observation time is indicated in minutes. Given that no source has been detected so far, a 99\% pre-trials upper limit above 100 GeV has been derived. The upper limit values are shown in the last two columns in units of integrated flux and as a percentage of the Crab nebula flux.}
\label{tab:uls}
\end{table*}

\section{Future work}

We plan to continue observations of the muons not covered so far in this work (i.e. UC1, 13, 14, 18 and 20) and to extend the program to include events detected by IceCube more recently that have a high probability (given their energy) of being astrophysical. The next steps will involve the analysis of \emph{Fermi}-LAT data at the muon locations, which is not as affected by EBL absorption as the VHE flux. 
In order to increase the sensitivity to transient sources, work is ongoing in IceCube to broadcast realtime alerts for interesting muons that will trigger follow-up observations with multi-wavelength instruments. We plan to participate in these follow-up observations when the alerts become publicly available.

%



\section{Acknowledgements}
\footnotesize{
VERITAS research is supported by grants from the U.S. Department of Energy Office of Science, the U.S. National Science Foundation and the Smithsonian Institution, and by NSERC in Canada. We acknowledge the excellent work of the technical support staff at the Fred Lawrence Whipple Observatory and at the collaborating institutions in the construction and operation of the instrument.
The VERITAS Collaboration is grateful to Trevor Weekes for his seminal contributions and leadership in the field of VHE gamma-ray astrophysics, which made this study possible. 

The IceCube Collaboration acknowledges the support from the following agencies: U.S. National Science Foundation-Office of Polar Programs, U.S. National Science Foundation-Physics Division, University of Wisconsin Alumni Research Foundation, the Grid Laboratory Of Wisconsin (GLOW) grid infrastructure at the University of Wisconsin - Madison, the Open Science Grid (OSG) grid infrastructure; U.S. Department of Energy, and National Energy Research Scientific Computing Center, the Louisiana Optical Network Initiative (LONI) grid computing resources; Natural Sciences and Engineering Research Council of Canada, WestGrid and Compute/Calcul Canada; Swedish Research Council, Swedish Polar Research Secretariat, Swedish National Infrastructure for Computing (SNIC), and Knut and Alice Wallenberg Foundation, Sweden; German Ministry for Education and Research (BMBF), Deutsche Forschungsgemeinschaft (DFG), Helmholtz Alliance for Astroparticle Physics (HAP), Research Department of Plasmas with Complex Interactions (Bochum), Germany; Fund for Scientific Research (FNRS-FWO), FWO Odysseus programme, Flanders Institute to encourage scientific and technological research in industry (IWT), Belgian Federal Science Policy Office (Belspo); University of Oxford, United Kingdom; Marsden Fund, New Zealand; Australian Research Council; Japan Society for Promotion of Science (JSPS); the Swiss National Science Foundation (SNSF), Switzerland; National Research Foundation of Korea (NRF); Danish National Research Foundation, Denmark (DNRF).}

\end{document}